# Token Ring Project

**Assist.Prof. Virgiliu Streian**
"Tibiscus" University of Timisoara, Romania
**University Assistant Adela Ionescu**
"Tibiscus" University of Timisoara, Romania

REZUMAT. Topologia ring este o configurare simplă utilizată pentru conectarea de procese ce comunică între ele. O serie de standarde de rețea ca de exemplu token ring, token bus, sau FDDI sunt bazate pe conectivitatea ring. În acest articol se va dezvolta o implementare a unui ring de procese care comunică între ele via pipe-uri. Procesele reprezintă nodurile din ring. Fiecare proces citește din intrarea lui standard și scrie în ieșirea lui standard. Procesul n-1 redirectează ieșirea lui standard către intrarea standard a procesului n printr-un pipe. În momentul în care structura ring-ului este concepută, proiectul poate fi extins la simularea rețelelor sau implementarea algoritmilor pentru excluziunea mutuală.

## 1 Single Process Ring

The following code segment connects the standard output of a process to its standard input through a pipe. Error checking has been omitted for simplicity.

```
#include<unistd.h>
int fd[2];
pipe(fd);
// X
dup2(fd[0],STDIN_FILENO);
dup2(fd[1],STDOUT_FILENO);
// XX
```

85

```
    close(fd[0]);
    close(fd[1]);
    // XXX
```

The following figure shows the file descriptor table after pipe A has been created (X). File descriptor entries [3] and [4] point to system file table entries that were created by a pipe call. A program can write to the pipe at this point by using a file descriptor value of 4 in a write.

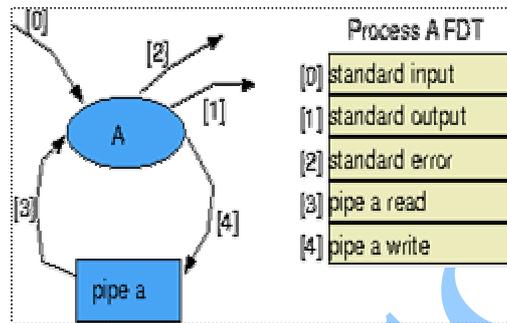

This figure shows the status of the file descriptor table after the dup2 commands (XX). At this point the program can write to the pipe using either 1 or 4 as the file descriptor values.

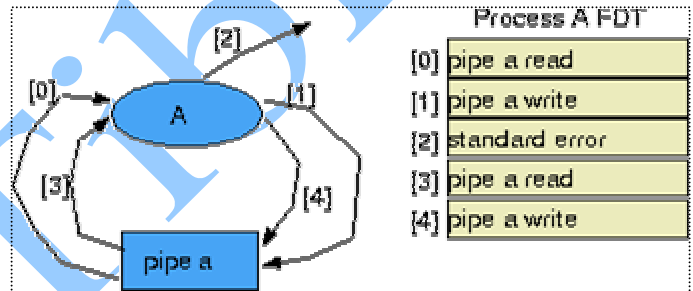

This figure shows the status of the file descriptor table after descriptors 3 and 4 are closed (XXX).

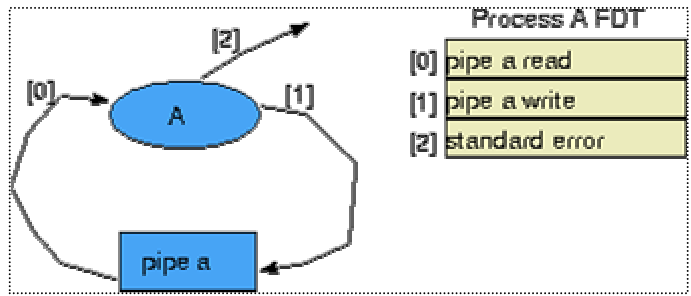

## 2  Two Process Token Ring

The following code segment creates a ring of two processes

```
#include <unistd.h>
int fd[2];
pid_t haschild;

pipe(fd);

dup2(fd[0], STDIN_FILENO);
dup2(fd[1], STDOUT_FILENO);

close(fd[0]);
close(fd[1]);

pipe(fd);

if (haschild = fork())
dup2(fd[1], STDOUT_FILENO); /* parent redirects std output */
else
dup2(fd[0], STDIN_FILENO);  /* child redirects std input */
close(fd[0]);
close(fd[1]);
```

The parent process redirects standard input through the first pipe to the standard output of the child and redirects standard output through the second pipe to the standard input of the child. The attached sequence of figures illustrate the connection mechanism.

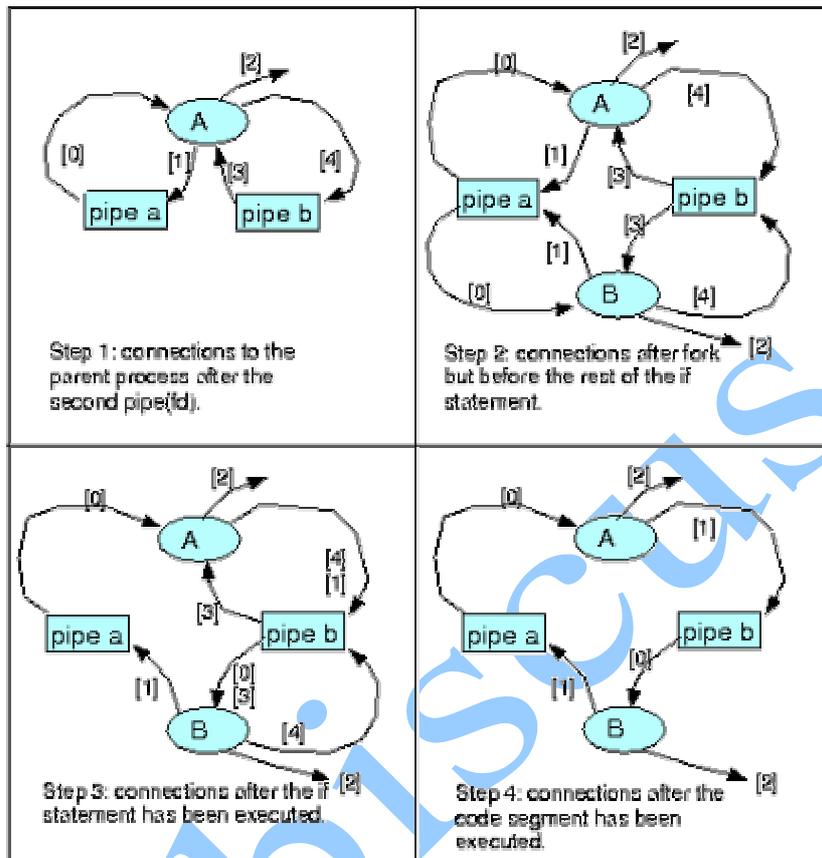

## 3  n Process Token Ring

The following code segment creates a ring of n processes. This is a sample C program for generating a unidirectional ring of processes. Invoke this program with a command-line argument indicating the number of processes on the ring. Communication is done via pipes that connect the standard output of a process to the standard input of its successor on the ring. After the ring is created, each process identifies itself with its process ID and the process ID of its parent.

/*  Un program care creaza un ring unidirectional intre n procese */
/*
   Un program simplu C de generare a unui ring unidirectional intre N procese. Startarea programului se face printr-o linie de forma nume_program N,  unde N este numarul de procese in ring.  Communicara e realizata via pipe-uri care conecteaza standard output al procesului la standard input succesorului sau din ring. Dupa ce s-a creat ring-ul, fiecare proces se autoidentifica prin ProcessID si ParentProcessID. In final fiecare proces se termina normal printr-un exit(0).
*/

```
#include <stdio.h>
#include <stdlib.h>
#include <unistd.h>
#include <string.h>
#include <errno.h>
#include <sys/wait.h>

int main(int argc,  char *argv[ ])
{
  int   i;          /* numarul procesului (starteaza de la 1)       */
  int   childpid;      /* indica un PID-ul copilului creat de un parinte  */
  int   pid;        /* indica procesului care s-a terminat       */
  int   nprocs;       /* numarul total de procese in ring (din linia de comanda) */
  int   fd[2];       /* tabela descriptorului de fisiere returnat de  pipe */
  int   status;       /* memoreaza starea dupa waitpid             */

   /* control pentru ca linia de comanda sa contina un numar valid de procese ce vor fi generate */
  if ( (argc != 2) || ((nprocs = atoi (argv[1])) <= 0) ) {
    fprintf (stderr, "Utilizare: %s nprocs\n", argv[0]);
    exit(1);
  }

  pipe (fd);     /* conecteaza std input la std output via un pipe */
  dup2(fd[0], STDIN_FILENO);
  dup2(fd[1], STDOUT_FILENO);
  close(fd[0]);
  close(fd[1]);
     /* creaza restul proceselor cu acest pipe de conectare */
  for (i = 1; i < nprocs;  i++) {
    pipe (fd);
    childpid = fork();    /* creaza un child */
```

```
        if (childpid > 0){
             /* pentru procesele parent se reasigneaza stdout */
           dup2(fd[1], STDOUT_FILENO);}
        else {
              /* pentru procesele child se reasigneaza stdin */
           dup2(fd[0], STDIN_FILENO);}

        close(fd[0]);
        close(fd[1]);
        if (childpid)       /* daca parent atunci iesi din for*/
           break;
      }; // end for
    /* Se introduce un waitpid inaintea lui fprintf final */
    while ((( pid = wait(&status)) > 0)) { // astept terminarea unui proces
     fprintf(stderr, "Inca un copil mort PID = %ld.\n", pid);
    }  // end while
                       /* zice ceva la iesire */
    fprintf(stderr,"Procesul[%ld],  ProcessID = %ld, ParentID = %ld\n",
       i, (int)getpid(), (int)getppid());
    exit (0);
  }    /* sfarsit main program */
```

## References

[Jur01]   **Ioan Jurca**, *Programarea Reţelelor de Calculatoare*, Editura de Vest Timişoara, 2001

[LS03]    **Lucian Luca şi Virgiliu Streian**, *Sistemul de operare Solaris*, Editura Mirton Timişoara, 2003

[LS05]    **Lucian Luca şi Virgiliu Streian**, *UNIX*, Editura Mirton Timişoara, 2005.

[RR96]    **K. Robbins and S. Robbins**, *Practical UNIX Programming: a guide to concurrency, communication, and multithreading*, Prentice-Hall, 1996

[SL98]    **Virgiliu Streian şi Lucian Luca**, *Sistemul de operare UNIX*, Editura Mirton Timişoara, 1998